\documentclass[twocolumn]{aastex61}
\usepackage{epstopdf}
\usepackage{soul}

\accepted{March 27, 2017}

\submitjournal{AJ}

\shorttitle{Chariklo and its Rings}
\shortauthors{Wood et al.}

\begin{document}

\title{THE DYNAMICAL HISTORY OF CHARIKLO AND ITS RINGS}

\correspondingauthor{Jeremy Wood}
\email{jeremy.wood@kctcs.edu}

\author{Jeremy Wood}
\affil{Hazard Community and Technical College, Community College Drive Hazard, KY USA 41701}
\affil{Computational Engineering and Science Research Centre, University of Southern Queensland West St, Toowoomba, QLD 4350, Australia}

\author{Jonti Horner}
\affil{Computational Engineering and Science Research Centre, University of Southern Queensland West St, Toowoomba, QLD 4350, Australia}
\affil{Australian Centre for Astrobiology, UNSW Australia, Sydney, NSW 2052, Australia}

\author{Tobias C. Hinse}
\affil{Korea Astronomy and Space Science Institute, 776 Daedukdae-ro, Yuseong-gu, Daejeon 305-348, Republic of Korea}
\affil{Armagh Observatory, College Hill, Armagh BT61 9DG, UK}

\author{Stephen C. Marsden}
\affil{Computational Engineering and Science Research Centre, University of Southern Queensland West St, Toowoomba, QLD 4350, Australia}

\newcommand{\change}[1]{\textcolor{red}{\textbf{#1}}}
\newcommand{\changeblue}[1]{\textcolor{blue}{\textbf{#1}}}

\begin{abstract}
Chariklo is the only small Solar system body confirmed to have rings. Given the instability of its orbit, the presence of rings is surprising, and their origin remains poorly understood. In this work, we study the dynamical history of the Chariklo system by integrating almost 36,000 Chariklo clones backwards in time for one Gyr under the influence of the Sun and the four giant planets. By recording all close encounters between the clones and planets, we investigate the likelihood that Chariklo's rings could have survived since its capture to the Centaur population. Our results reveal that Chariklo's orbit occupies a region of stable chaos, resulting in its orbit being marginally more stable than those of the other Centaurs. Despite this, we find that it was most likely captured to the Centaur population within the last 20 Myr, and that its orbital evolution has been continually punctuated by regular close encounters with the giant planets. The great majority ($>$ 99\%) of those encounters within one Hill radius of the planet have only a small effect on the rings. We conclude that close encounters with giant planets have not had a significant effect on the ring structure. Encounters within the Roche limit of the giant planets are rare, making ring creation through tidal disruption unlikely. 
\end{abstract}


\section{INTRODUCTION}

The Centaurs are a dynamically unstable population of small bodies in the outer Solar system. The first Centaur to be discovered, Chiron, was discovered in 1977. After the discovery, astronomers searched through archival images, revealing the presence of Chiron on old photographic plates, which allowed the object's orbit to be precisely determined. It was soon realised that Chiron followed an unusual path around the Sun, spending the vast majority of its time between the orbits of Saturn and Uranus \citep{KowaiCT:1979}. In the decades since Chiron's discovery, many other Centaurs have been found, all following unstable orbits in the outer Solar system. Though the definition of Centaur varies within the astronomical community, we will use the definition adopted by the Minor Planet Center that Centaurs move on orbits with semi-major axes between those of Jupiter and Neptune, and have perihelia beyond Jupiter's orbit\footnote{http://www.minorplanetcenter.net/iau/lists/Unusual.html  (accessed 15th January 2016)} \citep[e.g.][]{JewittD:2009, SheppardS:2000}. They exhibit extreme dynamical instability \citep[e.g.][]{HornerJ:2004a, BaileyBL:2009}, being scattered chaotically as a result of regular close encounters with the giant planets.

As a result of their extreme dynamical instability, the observed Centaurs can not simply be the last remaining members of a once larger, primordial population. Instead, the must be continually replenished. Over the years, a number of other Solar system small body populations have been suggested as potential sources for the Centaurs, including captured Oort Cloud comets (Emel'yanenko et al. 2005) \citep[e.g.][]{BrasserR:2012, FouchardM:2014}, the Jovian Trojans \citep[e.g.][]{HornerJ:2006, HornerJ:2012b}, and the Neptune Trojans \citep{HornerJ:2010a, HornerJ:2010b, HornerJ:2012a}.

The primary source population, however, seems likely to be the trans-Neptunian objects - principally the Scattered Disk \citep[e.g.][]{DiSistoRP:2007, VolkK:2008}, with a small contribution from the classical Edgeworth-Kuiper belt \citep[e.g.][]{LevisonHF:1997}. In turn, the Centaurs are thought to be the primary parent population for the short-period comets - with up to a third of Centaurs likely to be captured to that population at some point during their chaotic evolution \citep[e.g.][]{HornerJ:2004a}.

The largest known Centaur is Chariklo, with an estimated diameter of approximately 250 km\citep{FornasierS:2014}. It moves on a moderately eccentric orbit between the orbits of Saturn and Uranus, with a semi-major axis of 15.8 au. Early dynamical studies showed that Chariklo moves on a relatively stable orbit for a Centaur, with an estimated dynamical half-life of 10.3 Myr \citep{HornerJ:2004a}.

In 2013, observations of a chance stellar occultation by Chariklo revealed the unexpected presence of two narrow rings with radii 391 km and 405 km, respectively - making it the only small body in the Solar system confirmed to possess rings \citep{Braga-RibasF:2014}.

The discovery of Chariklo's rings was a great surprise and has prompted significant discussion on their nature and origin, as well as opening up the possibility that other small bodies such as Chiron could also possess rings \citep[e.g.][]{OrtizJL:2015, PanM:2016}.

A variety of mechanisms have been proposed to explain the rings, including leftover debris from a collision with another small body, debris from the tidal disruption of another small body \citep{ElMoutamidM:2014}, partial tidal disruption of Chariklo itself \citep{HyodoR:2016} and dust particles sent into orbit due to an outflow of CO and/or N$_2$ from Chariklo as a result of cometary activity \citep{PanM:2016}.

\citet{ElMoutamidM:2014} suggest the possibility that shepherd satellites could exist around Chariklo making the rings more stable. Such satellites are known to sculpt the rings of the giant planets - with several examples found in the Saturnian system alone \citep[e.g.][]{2009sfch.book..375C}.

Whilst such shepherding satellites have not yet been found in orbit around Chariklo, their presence would potentially ensure the long term survival of the ring system.

The presence of rings around Chariklo is perhaps particularly surprising when one considers that the orbits of Centaurs are highly chaotic, as a result of the gravitational influence of the giant planets \citep{TiscarenoM:2003, BaileyBL:2009}. On average, a Centaur remains just 10 Myr in the Centaur region \citep{LevisonHF:1994, DonesL:1996, TiscarenoM:2003, HornerJ:2004a} which is far less than the age of the Solar system (4.6 Gyr).

During their lifetime, Centaurs cross the orbits of the giant planets and most likely experience multiple close encounters within one Hill radius of those planets during their stay in the Centaur region \citep{TiscarenoM:2003, BaileyBL:2009, AraujoRAN:2016}. This opens up the possibility that Chariklo has had a close encounter with a giant planet at some time in its past which was so close that the rings as they exist today would not have survived.

The goal of this work is to determine the dynamical history of Chariklo and its rings; and to examine the chaoticity and lifetimes of Chariklo-like orbits in semi-major axis-eccentricity space. In section two, we present the known properties of Chariklo based on earlier observational work and introduce criteria which we will apply to measure the severity of a close encounter. In section three we describe our methodology, before presenting our results in section four, and summarising our conclusions in section five.

\section{CHARIKLO PROPERTIES AND THEORY}

Chariklo was discovered in 1997 by the Spacewatch program\footnote{http://spacewatch.lpl.arizona.edu\/discovery.html accessed 29th October, 2016}, moving on an orbit that lies between those of Saturn and Uranus, within 0.09 au of the location of the 4:3 mean motion resonance with Uranus. Its physical properties and those of its rings are presented in Table~\ref{chariklo_sizes}. Orbital elements of Chariklo are shown in Table~\ref{chariklo_orbit}.

Since its discovery, a number of groups have carried out observations of Chariklo at a variety of wavelengths, with the goal of disentangling its surface composition. Despite the work that has been carried out, there remains significant disagreement on the Centaur's surface composition. \citet{GroussinO:2004} report that the reflectance spectrum of Chariklo is consistent with a surface composed of 80\% refractory material and 20\% water ice.

Guilbert et al. (2009) reported water ice in the combined spectrum of Chariklo+rings and Duffard et al. (2014) showed that the water ice feature comes only from the rings, and not from Chariklo. The rings are believed to be composed of water ice, silicates, tholins and some amorphous carbon (Duffard et al. 2014). 

To date, no cometary activity has been detected for Chariklo, despite it passing through perihelion in the last decade. However, this does not rule out the possibility that it may have displayed cometary activity in the past \citep{GuilbertA:2009}.

Backwards integrations show that Chariklo has a backward half-life of 9.38 Myr, some 1.6 Myr longer than the next largest Centaur Chiron \citep{HornerJ:2004b}.


\subsection{The Stability of Rings Through Close Encounters: The ``Ring Limit" Criterion}


\begin{table*}
\begin{center}
\caption{Properties of Chariklo and its rings. [1] \citet{AltenhoffWJ:2001} [2] \citet{JewittD:1998} [3] \citet{GroussinO:2004} [4] \citet{Braga-RibasF:2014} [5] \citet{ElMoutamidM:2014} [6] \citet{FornasierS:2014} [7] \citet{CampinsH:2000} [8] \citet{DuffardR:2014} [9] \citet{BrownME:1998}.}\label{chariklo_sizes}
\begin{tabular}{|c|c|c|}
\hline
Property&Value&Uncertainty\\
\hline
Radius (km)&137$^{1}$&10\\
&151$^{2}$&15\\
&118$^{3}$&6\\
&124$^{4}$&9\\
&125$^{5}$&\\
&119$^{6}$&5\\
&&\\
Albedo&0.045$^{7}$&0.01\\
Composition&amorphous carbon 60\%& \\
&silicates 30\%&\\
&organics 10\%$^{8}$&\\
&$3\%$ water ice$^{9}$&\\
&&\\
Inner Ring Width (km)& 7$^{4}$&\\
Inner Ring Radius (km)&391$^{4}$&\\
Outer Ring Width (km)&3$^{4}$&\\
Outer Ring Radius (km)&405$^{4}$&\\
&&\\
Ring Composition&20\% water ice&\\
&40-70\% silicates& \\
&10-30\% tholins& \\
&small quantities of amorphous carbon$^{8}$&\\
\hline
\end{tabular}
\end{center}
\end{table*}

\begin{table}
\begin{center}
\caption{Orbital elements of Chariklo taken from the Asteroids Dynamic WWW\footnote{http://hamilton.dm.unipi.it/astdys/; accessed 31st Dec., 2015} site for epoch MJD 2,457,600.0 based on an observational arc of 9,684.35 days.}
\label{chariklo_orbit}
\begin{tabular}{|c|c|}
\hline
Element&Value $\pm$ Uncertainty (1-sigma)\\ 
\hline
eccentricity&0.172265 $\pm$ 1.8036e-06\\
semi-major axis&15.77739 $\pm$ 3.75e-05 au\\
inclination&23.408508 $\pm$ 9.5473e-06 deg.\\
longitude of ascending node&300.38512 $\pm$ 2.9189e-05 deg.\\
longitude of perihelion&241.9872 $\pm$ 0.00014188 deg.\\
Mean anomaly&65.9988 $\pm$ 0.00029106 deg.\\
\hline
\end{tabular}
\end{center}
\end{table}
The severity of a close encounter between a small body (such as Chariklo) and one of the giant planets has been shown to depend on the closest approach distance of the encounter, and the velocity of the small body at infinity. \citep{AraujoRAN:2008, HyodoR:2016}. In order to determine the dynamical history of Chariklo and its rings, we neglect velocity effects following \citet{AraujoRAN:2016} and compare the minimum close encounter distance between Chariklo and a planet to three different critical distances within the Hill sphere of the planet. The first of these is the distance between Chariklo and a planet at which tidal forces can disrupt a Chariklo-ring particle binary pair instantaneously. This tidal disruption distance, $R_{td}$, for a binary consisting of a massless, outermost ring particle in a circular orbit and Chariklo is given by:

\begin{equation}
R_{td}\approx r\Big(\frac{3M_p}{m_{ch}}\Big)^{\frac{1}{3}}\label{tidal_disrupteqn}
\end{equation}

\noindent where $M_p$ is the mass of the planet, $m_{ch}$ is the mass of Chariklo and $r$ is the orbital radius of a ring particle \citep{AgnorCB:2006, PhilpottCM:2010}. When Chariklo is just within the tidal disruption distance to a planet, an outermost ring particle is just outside Chariklo's Hill sphere. According to \citet{AraujoRAN:2016} the minimum distance obtained between Chariklo and a planet during a close encounter must be $\le 10R_{td}$ in order for the encounter to have a significant effect on the rings. We will refer to this distance as the 'ring limit', $R$. They considered the effect `noticeable' if the maximum change in eccentricity of any orbiting ring particle was at least 0.01.

But there is one more critical distance to consider. At an even closer distance to a planet is the Roche Limit - the distance within which a small body like Chariklo can be torn apart by tidal forces. For a small, spherical satellite of a planet, the equation for the Roche limit is \citep{MurrayCD:1999}:

\begin{equation}
R_{roche}\approx R_{ch}\Big(\frac{3M_p}{m_{ch}}\Big)^{\frac{1}{3}}\label{rocheeqn}
\end{equation}
\noindent where $R_{ch}$ is the physical radius of Chariklo.

Since closer approaches have a larger effect than more distant ones, the minimum distance, $d_{min}$, obtained between Chariklo and a planet during a close encounter can be used to assess severity. \par
We now present in Table~\ref{CE_severity} a severity scale based on $d_{min}$ relative to the distances $R_H,R,R_{td}$ and $R_{roche}$.\par

\begin{table} [h]
\begin{tabular} {|c|c|}
\hline
Minimum Distance Range&Severity\\
\hline
$d_{min}\ge R_H$&Very Low\\
$10 R_{td}\le d_{min} < R_H$&Low\\
$R_{td}\le d_{min}<$ $10R_{td}$&Moderate\\
$R_{roche}\le d_{min}< R_{td}$&Severe\\
$d_{min}<R_{roche}$&Extreme\\
\hline
\end{tabular}\caption{A scale ranking the close encounter severity between a ringed small body and a planet based on the minimum distance obtained between the small body and the planet, $d_{min}$, during the close encounter. $R_H$,$10R_{td}$},$R_{td}$ and $R_{roche}$ are the Hill radius of the planet, $R=$ 10 $\times$ tidal disruption distance, tidal disruption distance and Roche limit respectively (see text for details).\label{CE_severity}
\end{table}

\section{METHOD}

\subsection{Chariklo}

In order to determine whether Chariklo has experienced sufficiently close encounters with the giant planets to disrupt its rings during its life, we need to be able to determine its historical dynamical evolution.

To do this, we follow the same methodology as that used in previous studies of dynamically unstable objects \citep[e.g.][]{HornerJ:2004a, HornerJ:2010a, KissCS:2013, PalA:2015} and follow the evolution of a suite of clones of Chariklo backwards in time for a period of 1 Gyr. By following the evolution of a large population of Chariklo clones, we can obtain a statistical overview of the object's potential past history.

As in those earlier works, we created a grid of test particles, centered on the best-fit solution taken from Table~\ref{chariklo_orbit}, by incrementing the semi-major axis, $a$, eccentricity, $e$, and inclination, $i$, of the test particles in even steps through the full $\pm 3\sigma$ uncertainty ranges in those elements. We held the three rotational orbital elements, argument of perihelion, longitude of ascending node and Mean Anomaly constant across our population of clones.

33 massless test particles per orbital parameter were created for parameters $a, e$ and $i$ to yield a total of $33^3 =$ 35,937 test particles. Test particles were evenly spaced across the full uncertainty range of the orbital parameter.

The initial orbital elements of the four giant planets were found using the NASA JPL HORIZON ephemeris\footnote{http://ssd.jpl.nasa.gov/horizons.cgi?s\_body=1$\#$top (accessed 31st December 2015)} for epoch Jan 1, 2000 at UT 00:00. Inclinations and longitudes for both Chariklo and the planets were relative to the ecliptic plane.

The planets were then integrated (within the heliocentric frame) to the epoch MJD 2,457,600.0 - the epoch of the Chariklo clones using the \textit{Hybrid} integrator within the \textsc{Mercury} N-body dynamics package \citep{ChambersJE:1999}. Test particles and planets were then integrated backwards in time for 1 Gyr in the 6-body problem (Sun, four giant planets and test particle) subject only to the gravitational forces of the Sun and giant planets. This integration time is 100 times longer than the typical lifetime of a Centaur ($\sim$ 10 Myr). Therefore, the conclusions presented in this study are limited to within this time span.

For the symplectic integration we chose a time step of 40 days \citep{HornerJ:2004a, HornerJ:2004b} corresponding to approximately 1\% of the orbital period of Jupiter, the innermost planet at the start of our integrations ensuring an accurate orbit calculation for the giant planets and the particle during non-close encounter epochs (e.g.~Tiscareno \& Malhotra 2003).

We set the accuracy tolerance parameter for the switch-over integration algorithm to be $10^{-12}$. This ensured an accurate integration of the test particle during epochs of high eccentricity excursions as a result of close encounters. A close encounter was said to have occurred when the distance between a test particle and a planet was $\le$ 3 Hill radii. The time of every close encounter between a test particle and any planet was recorded along with the instantaneous planet and test particle $a-e-i$ elements and the minimum separation obtained between the test particle and planet, $d_{min}$.

Test particles were removed from the simulation by colliding with a planet, upon reaching a barycentric distance of 1,000 au, achieving $e\ge1$ or by approaching within 0.005 au of the Sun. Removal times were recorded.

Moving backwards in time, the number of test particles in the Centaur region was assumed to decrease exponentially over some time interval according to the standard radioactive decay equation \citep{HornerJ:2004b}:

\begin{equation}
N=N_oe^{-\lambda t}
\label{radioactive_decay}
\end{equation}

Here, $N_o$ is the initial number of test particles, $N$ is the number of test particles remaining in the Centaur region at a time $t$ and $\lambda$ is the decay constant. The decay constant can be found from the slope of the best-fit line of a graph of  ln($\frac{N}{N_o}$) vs time. Then the half-life, $\tau$, is given by:

\begin{equation}
\tau=\frac{\textnormal{-ln}(0.5)}{\lambda}
\label{half_life_equation}
\end{equation}

The data for the number of test particles remaining in the Centaur region at a time $t$ was fit to Equation~\ref{radioactive_decay} to obtain the decay constant. Then Equation~\ref{half_life_equation} was used to find the half-life.

The half-life gives a best first estimate to Chariklo's age as a Centaur - with 50\% of the clones of Chariklo being ejected within that time period. We also used the half-life in Equation~\ref{radioactive_decay} to determine the time at which 99\% of all Chariklo-like objects would have left the Centaur region.

\subsection{The Severity of Close Encounters and the Mass of Chariklo}

In order to gain an understanding of whether Chariklo's rings existed prior to its capture to the Centaur region, or are a more recent addition, we can investigate the times at which test particles had encounters with the planets that were sufficiently close to disrupt the rings.

If the great majority of clones were to experience even a few severe or greater disruptive encounters or a large number of low to moderate encounters, this would suggest that Chariklo's rings most likely formed in the relatively recent past.
 
On the other hand, if relatively few clones have encounters deep enough to disturb the rings, then it is clearly feasible that the rings could be primordial (and, equally, such infrequent close encounters might in turn suggest that any origin for the rings involving the tidal disruption of Chariklo or an ancient satellite seems unlikely).
 
We therefore examined the depths and timings of the close encounters between test particles and planets and ranked the severity of each encounter using the scale in Table~\ref{CE_severity}.

As Table~\ref{CE_severity} along with equations~\ref{tidal_disrupteqn}, and~\ref{rocheeqn} show, the severity of a close encounter depends in part on Chariklo's mass. This mass was estimated using the average density of Chariklo from \citet{Braga-RibasF:2014} of 1,000 $\frac{\text{kg}}{\text{m}^3}$ and the radius value of 125 km from \citet{ElMoutamidM:2014}. A mass of 8.18 $\times 10^{18}$ kg was obtained. This calculation assumed that the shape of Chariklo was a perfect sphere, as it is nearly spherical with the major to minor axis ratio of 1.1 \citep{FornasierS:2014}.

\subsection{MEGNO and Lifetime maps}

In addition to our N-body integrations of Chariklo's orbital evolution, a complementary suite of calculations were performed to examine the wider dynamical context of Chariklo's orbit.

Since sampling very large regions of phase space is impractical with full-scale N-body integrations, we instead generated a MEGNO (Mean Exponential Growth factor of Nearby Orbits; \citealt{CincottaPM:2003}) map for the region of phase space bound by $14  \textnormal{ au} \le a \le 19  \textnormal{ au}$ and $e \le 0.8$ for Chariklo-like orbits. These are orbits which initially have the same orbital parameters as Chariklo except for semi-major axis and eccentricity.

The resolution of the map was 1024 $\times$ 800 pixels. The map was constructed by integrating one test particle per pixel or 30,000 test particles total using the Gragg-Bulirsh-Stoer \citep{HairerE:1993} method.

The initial values of $a$ and $e$ for each test particle were determined by each location of a pixel on the map. The integration algorithm makes use of a variable step-size determined by a relative and absolute tolerance parameter which both were set to be close to the machine precision. The total integration time for each particle in the a-e grid was 1 Myr.

MEGNO maps show the chaoticity of a region of $a-e$ space by calculating a parameter $\langle Y\rangle$ which is proportional to the Lyapunov characteristic exponent at each point. The reader is referred to \citet{CincottaPM:2000}, \citet{GozdziewskiK:2001}, \citet{CincottaPM:2003}, \citet{GiordanoCM:2004} and \citet{HinseTC:2010} for more details on MEGNO maps. For an explanation of  Lyapunov characteristic exponents the reader is referred to \citet{WhippleAL:1995}.

$\langle Y\rangle$ will asymptotically converge towards 2.0 for quasi-periodic orbits and diverge from 2.0 for chaotic orbits as the system is allowed to evolve in time.

For this work, quasi-periodic orbits were color coded blue and highly chaotic orbits yellow. Test particles were removed by following the same criteria as for the long-term integration described earlier in this work. In addition, we terminated a given integration when $\langle Y\rangle$ $> 12$ which indicates a strong degree of chaos.

When a test particle was removed, the time of removal and the $\langle Y\rangle$ value were recorded. If a test particle survived the entire simulation then its removal time was recorded as 1 Myr. A lifetime map was then generated in conjunction with the MEGNO map covering the same a-e grid space. In the life-time map shortest removal times were color coded black and the longest with yellow.

\section{RESULTS}

\subsection{The Dynamical History of Chariklo}

Over 70 million close encounters within 3 Hill radii were recorded, with roughly 7.1 million of these being at distances less than one Hill radius. The close encounters were analysed using eight different subsets of the entire encounter dataset. Five of those subsets examined close encounters whilst the clone in question was a member of one of the Solar system's various small body populations (as detailed below), with the other three described as follows:

\begin{enumerate}
\item The set of first close encounters - a first close encounter is the earliest time chronologically at which a close encounter occurred. Each test particle had one and only one of these.
\item The set of close encounters at any time at which the test particle was classified as a Centaur. Each test particle had more than one of these.
\item The set of earliest close encounters chronologically (not necessarily a first close encounter) at which each test particle was classified as a Centaur. Each test particle had one and only one of these.
\end{enumerate}

The subsets of close encounters based on the membership of the clone in a small body population when the close encounter occurred are described as follows:
\begin{enumerate}
\item  Inner SS - $a\le a_J$
\item Comet - $a>a_J$ and $q<a_J$
\item Centaur - $a_J<a<a_N$ and $q>a_J$
\item TNO - $a\ge a_N$
\item Ejection - the test particle was being ejected from the Solar system at the time of the close encounter
\end{enumerate}
Where $a$ is the semi-major axis of the test particle at the time of the close encounter, $a_J$ the semi-major axis of Jupiter, $a_N$ the semi-major axis of Neptune and $q$ the perihelion distance of the test particle.

\begin{table}
\begin{center}
\caption{The percentage of close encounters as a function of membership of the different small body populations Chariklo's clones occupied through the course of the integrations.}\label{CE_region}
\begin{tabular} {|c|c|}
\hline
Region&Percent\\
\hline
Inner SS&7\\
Comet&9\\
Centaur&53\\
TNO&31\\
Ejection&0.4\\
\hline
\end{tabular}
\end{center}
\end{table}

The percentage of close encounters which occurred when the test particle was in each of the five population subsets is shown in Table~\ref{CE_region}. The Centaur and TNO subsets dominate with 53\% and 31\% of the close encounters respectively. This implies but does not prove that Chariklo entered the Centaur region from beyond Neptune.

To determine the dynamical history of Chariklo and if Chariklo did enter the Centaur region from beyond Neptune, three of the subsets were investigated.

First, analysis of the Centaur subset showed that the average time between consecutive close encounters within 3 Hill radii of a planet in the Centaur region was 8 kyr. Therefore, on Myr timescales, the earliest time chronologically of a close encounter of a test particle while classified as a Centaur was taken to be the approximate time of insertion of the test particle into the Centaur region.

The set of earliest close encounters chronologically in which each test particle was classified as a Centaur was used to determine the number of test particles in the Centaur region as function of time. This data was analyzed by fitting it to Equation~\ref{radioactive_decay} with time measured from the start date backwards in time.

\begin{figure}
\begin{center}
\includegraphics[width=\columnwidth]{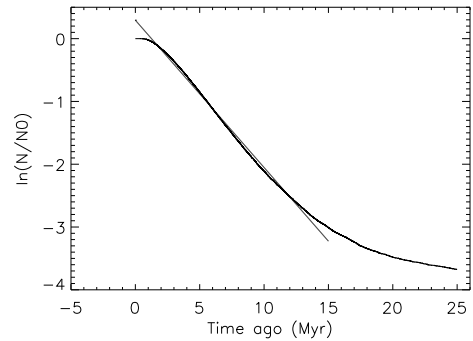}
\caption{The decay of test particles from the Centaur region moving backwards in time. The decay is exponential from 1.0 Myr to 14.1 Myr ago. The straight line is the line of best fit over this time interval. It has a slope of -0.2346 Myr$^{-1}$ and linear regression coefficient of -0.998. The slope was used to find the half-life of $\sim$ 3 Myr. At a time of 14.1 Myr ago only 5.57\% of the test particles were in the Centaur region. Note the backwards `s' shape. The bin size is 1 kyr.}
\label{ln_fraction_close_v_time_ago}
\end{center}
\end{figure}

Figure~\ref{ln_fraction_close_v_time_ago} shows the decay of the number of test particles in the Centaur region moving backwards in time. Note the reverse `s' shape of the graph. It took $\sim$1.0 Myr for the swarm of clones to disperse enough so that the decay could start to be exponential.

\begin{figure}
\begin{center}
\includegraphics[width=\columnwidth]{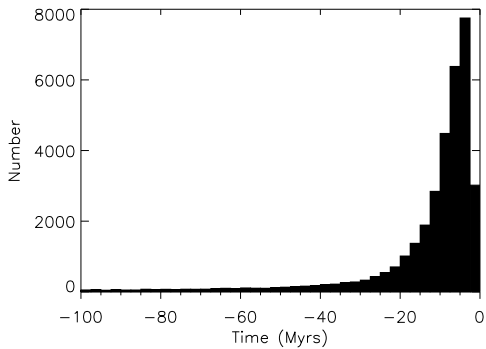}
\caption{A histogram of the number of first close encounters over the last 100 Myr. The bin size is 2.5 Myr.}
\label{CE_histogram_3_HR_cease_time}
\end{center}
\end{figure}

The exponential decay lasted from around 1.0 Myr to 14.1 Myr ago. The best-fit line over the interval is shown as the straight line. The linear regression coefficient of the line was -0.998 which shows a strong anti-correlation.

Between 14.1 Myr - 25 Myr ago the decay proceeded more slowly and no longer strongly correlated to the best-fit line. This occurred because by that time many remaining test particles had evolved onto more stable orbits, which in turn took longer to decay.

From the slope of the best-fit line of -0.2346 Myr$^{-1}$ and Equation~\ref{half_life_equation}, the half-life with respect to removal from the Centaur region was calculated to be $\sim$ 3 Myr. Since we expect half-lives with respect to removal from the Centaur region to be less than half-lives with respect to removal from the Solar system, this value is in broad agreement with the work of \citet{HornerJ:2004b} who found a backwards integrated half-life of Chariklo with respect to removal from the Solar system of 9.78 Myr.

Using the 3 Myr value for the half-life in Equation~\ref{radioactive_decay} suggests that there is a 99\% probability that Chariklo was injected into the Centaur region at some time within the last 20 Myr.
  
Finally, once a likely time frame for injection into the Centaur region was established, the set of first close encounters was studied to determine from what region of the Solar system Chariklo entered the Centaur region. Figure~\ref{CE_histogram_3_HR_cease_time} shows a histogram of the number of first close encounters over the last 100 Myr.

Table~\ref{first_ce} shows statistics on the first close encounters by region of the Solar system. From the table it seems most likely that Chariklo entered the Centaur region from an orbit outside that of Neptune, perhaps from the Edgeworth-Kuiper Belt \citep{HornerJ:2004b} or Scattered Disk \citep{DuncanM:2004, DiSistoRP:2007}. Two factors point to this conclusion:

\begin{enumerate}
\item The small percentage (2\%) of the subset of first close encounters which were also members of the inner Solar system subset makes it statistically unlikely that Chariklo was captured directly to the inner Solar system from elsewhere (such as a long-period comet orbit, or the main asteroid belt) and then migrated outwards to the Centaur population.
\item The much larger percentage (63\%) and earliest chronological mean time of the subset of first close encounters which were also members of the TNO subset makes it statistically likely that Chariklo was a TNO before becoming a Centaur.
\end{enumerate}

\begin{table}
\begin{center}
\caption{Statistics on the set of first close encounters by small body population of the Solar system.}\label{first_ce}
\begin{tabular}{|c|c|c|}
\hline
Subset&Percent&Mean Time Ago (Myr)\\
\hline
Inner SS&2  & 15.7\\
Comet  &21&12.1\\
Centaur&6&15.5\\
TNO&63&32.6\\
Ejection&8&31\\
\hline
\end{tabular}
\end{center}
\end{table}

These dynamical results potentially complement the observed physical properties of Chariklo, which also suggest both an origin beyond the orbit of Neptune, and that the object has not spent a protracted period in the inner Solar system.

First, the presence of volatiles on Chariklo's surface suggests that it has not spent lengthy periods interior to the Solar system's ice-line where most sublimation of volatile material occurs \citep{WhippleFL:1979, DISistoRP:2009, BrownJC:2011}.

Indeed, \citet{LevisonHF:1997} suggest that just 25 kyr in the inner Solar system is enough to entirely devolatilise comets. However, it should be noted that Chariklo is significantly larger than the nuclei of short-period comets  \citep{WeissmanPR:2008} - and so could have potentially contained far more volatile material, and would therefore been able to survive a longer period of devolatilisation. Still, the presence of volatiles does suggest an origin beyond the ice-line - and most likely, beyond the orbit of Neptune.

Nevertheless, though the percentages of close encounters occurring in the inner Solar system and Comet subsets are relatively small they are not negligible.

This allows for the possibility that Chariklo could have been active for brief periods in its past, and its rings replenished by cometary activity. The only caveat is that Chariklo would have needed to migrate outward from such an orbit before its volatiles were extinguished. However, such inward and outward migrations are dynamically feasible \citep{HornerJ:2004a}.

It should also be noted that for Chariklo to exhibit comet-like activity it would not be necessary for the orbit to be in the inner Solar system as Centaurs beyond Jupiter are known to be active \citep{JewittD:2009}.

To determine which planet dominated the close encounters in each population, the close encounters of each of the five population subsets were subdivided by planet. The results are shown in Table~\ref{CE_by_planet}.

Uranus dominated the number of close encounters of the Centaur subset followed by Saturn, Neptune and Jupiter. In the TNO subset, Neptune dominated followed by Uranus, Saturn, and Jupiter. Thus, statistically, Neptune is most likely responsible for perturbing Chariklo into the Centaur region over time.

Jupiter dominated the number of close encounters of the other three subsets - Inner SS, Comet and Ejection.

\begin{table}
\begin{center}
\caption{The number of close encounters within one Hill radius by planet and population. Most close encounters occurred between the Centaur population and Uranus.}\label{CE_by_planet}
\begin{tabular} {|c|c|c|c|c|}
\hline
Population&J&S&U&N\\
\hline
Inner SS&492255&1531&0&0\\
Comet&452415&139289&11879&6652\\
Centaur&56142&991571&2033886&717860\\
TNO&9026&224005&475861&1465906\\
Ejection&18687&9671&479&571\\
\hline
\end{tabular}
\end{center}
\end{table}

\subsection{The Dynamical History of the Rings of Chariklo}

The values of all ring limits, tidal disruption distances and Roche limits are shown by planet in Table~\ref{R_values}.

Jupiter had the largest value of $R$ at 0.02400 au and Uranus the smallest at 0.008584 au. All values of $R$ were well within 1 Hill radius of each planet by an order of magnitude or larger.

Thus, to have a close encounter of at least moderate severity, it must be far closer than the size of the planet in question's Hill sphere - sufficiently close, in fact, that it would be placed within the domain of the regular satellites of that planet.
 
For example, to have a moderate close encounter with Jupiter, Chariklo would have to approach the giant planet at a distance similar to the orbital radius of Themisto, or roughly a factor of four times more distant from the planet than Callisto.
 
In other words - disruptive encounters require very close encounters, and hence might be expected to be relatively infrequent.
 
This hypothesis is well supported by our data as can be seen in Tables~\ref{CE_severity2} and~\ref{severity_by_planet}. Every single clone of Chariklo experienced multiple close encounters, however, the great majority of these approaches were relatively distant. Over 99\% of all close encounters were outside the ring limit $R=$ $10~R_{td}$ for the planet in question. Therefore, we conclude that planetary close encounters have not played a major role in the disruption of rings.

\begin{table}
\begin{center}
\caption{The Hill radii, ring limits, tidal disruption distances and Roche limits for each giant planet, see text for details.}\label{R_values}
\begin{tabular}{|c|c|c|c|c|}
\hline
Planet&$R_H$ (au)&$R$ (au)&$R_{td}$ (au)&$R_{roche}$ (au)\\
\hline
J&0.3387&0.02400&0.002400&7.408$\times 10^{-4}$\\
S&0.4128&0.01606&0.001606&4.956$\times 10^{-4}$\\
U&0.4473&0.008584&0.0008584&2.649$\times 10^{-4}$\\
N&0.7704&0.009069&0.0009069&2.799$\times 10^{-4}$\\
\hline
\end{tabular}
\end{center}
\end{table}

\begin{table}
\begin{center}
\caption{The number and percent of close encounters by severity which occurred within one Hill radius of any planet.}\label{CE_severity2}
\begin{tabular} {|c|c|c|}
\hline
Severity&Number&Percent\\
\hline
Low&7084469&100.0\\
Moderate&21953&0.0\\
Severe&1025&0.0\\
Extreme&239&0.0\\
\hline
\end{tabular}
\end{center}
\end{table}

\begin{table}
\begin{center}
\caption{Severity of close encounters by planet which occurred within one Hill radius of any planet. 99\%, of close encounters were of low severity. Only 0.0034\% of close encounters were of extreme severity.}\label{severity_by_planet}
\begin{tabular}{|c|c|c|c|c|}
\hline
Severity&J&S&U&N\\
\hline
Low&1012998&1360893&2520339&2190239 \\
Moderate&14685&4884&1675&709 \\
Severe&707&219&71&28 \\
Extreme&135&71&20&13 \\
\hline
\end{tabular}
\end{center}
\end{table}

Just 35\% of the clones experienced at least one encounter within $10R_{td}$. Thus, over half of the clones never experienced even at least a moderate close encounter.

Furthermore, since only 0.0034\% of the close encounters were extreme, it is unlikely (but still possible) that the rings were created by gaseous outflow during a close encounter \citep{HyodoR:2016} because this would require Chariklo to be closer to the planet than its Roche limit.

This theory of ring formation may be further put in doubt if the purported rings around Chiron \citep{OrtizJL:2015} and the Saturnian satellites Rhea and Iapetus\citep{2016arXiv161203321S} are confirmed because it would suggest that rings around small bodies are more common and are not 
formed by a very rare extreme close encounter.

It should be noted that no age of the rings of Chariklo can be stated with absolute certainty since the total effects of gaseous outflow, shepherd satellites (if any), ring replenishment and non-gravitational forces are unknown.


\subsection{MEGNO and Lifetime maps}

\begin{figure}
\begin{center}
\includegraphics[width=\columnwidth]{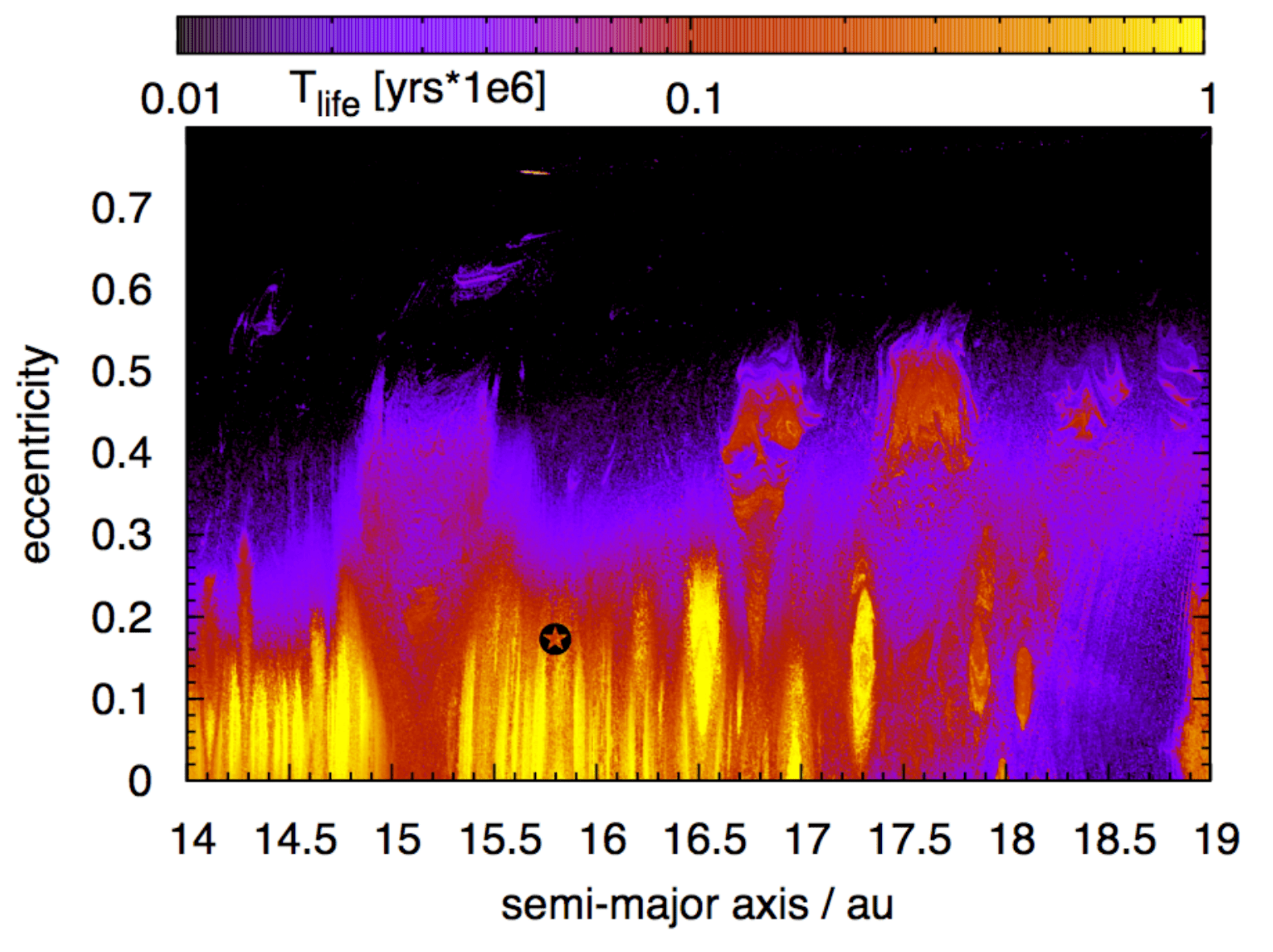}
\includegraphics[width=\columnwidth]{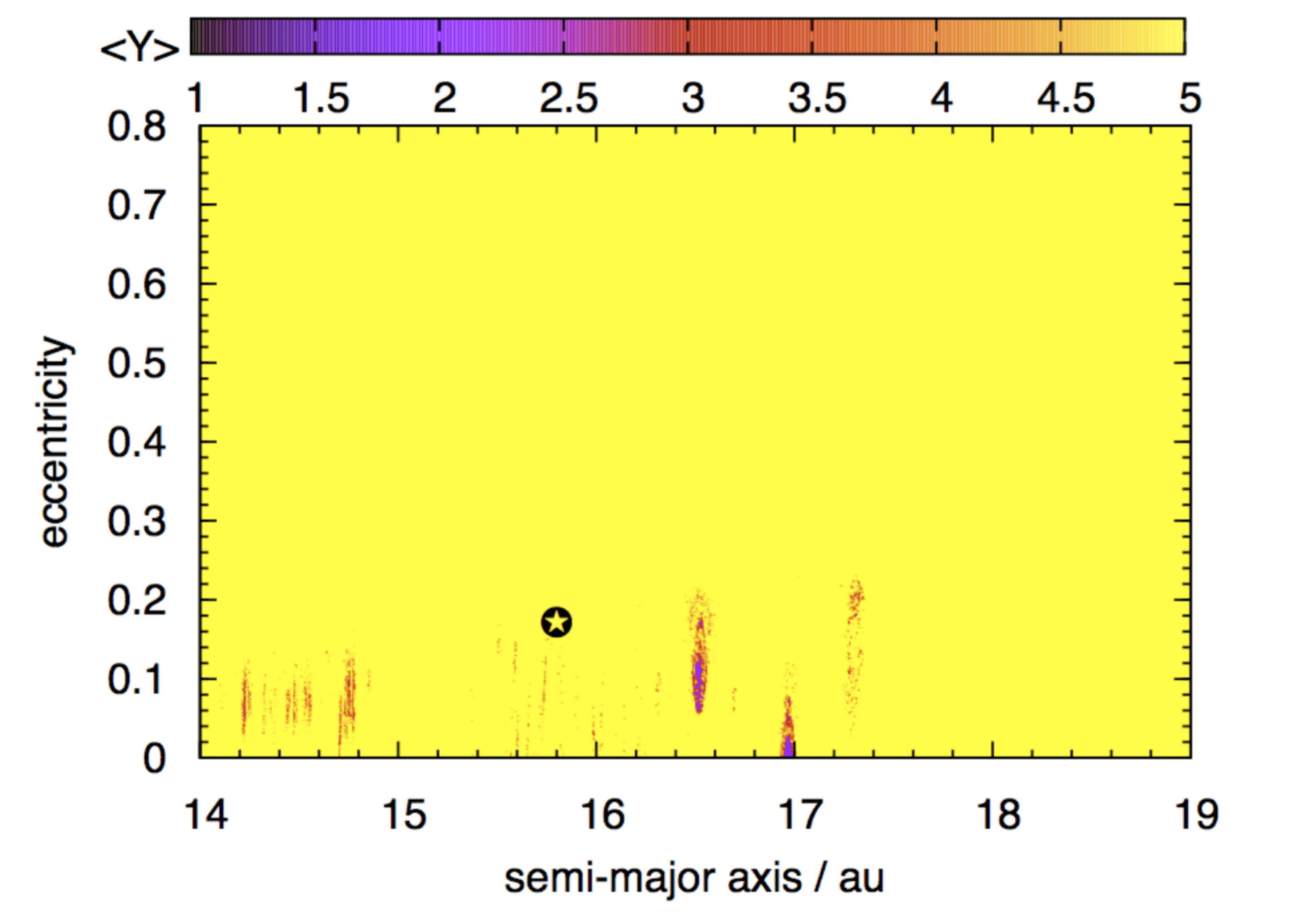}
\caption{The lifetime map (top panel) and the MEGNO map (bottom panel) of Cariklo-like orbits. Chariklo is located at $a=15.8$ au and $e=0.172$ and is marked by the star. For the top panel the longest lifetimes are shown in yellow and the shortest in black, while for the bottom panel highly chaotic orbits are shown in yellow and the least chaotic are blue.}\label{Chariklo_lifetime_MEGNO}
\end{center}
\end{figure}

The lifetime map in Figure~\ref{Chariklo_lifetime_MEGNO} shows that the longest lifetimes for Chariklo-like orbits in the $a-e$ region bound by $14\textnormal{ au}\le a \le 19\textnormal{ au}$ and $0\le e \le 0.8$ lie mostly in the rectangle bound by $14\textnormal{ au}\le a \le 17.4\textnormal{ au}$ and $0\le e < 0.26$. 

The effect of eccentricity on lifetime can clearly be seen. In the range $0.26 \le e \le 0.55$ virtually no lifetimes of 1 Myr can be seen for any value of $a$. Orbits with $e > \sim 0.55$ have noticeably shorter lifetimes for nearly all $a$ values to as low as 0.01 Myr. We attribute this drop in lifetime to the crossing of Saturn's orbit. 

The MEGNO map in Figure~\ref{Chariklo_lifetime_MEGNO} shows that the entire region is dominated by highly chaotic orbits ($\langle Y\rangle$ $\ge 5$).

Unlike lifetime, chaoticity does not display a clear relationship with eccentricity for constant $a$. Instead, relatively small islands of low chaoticity (quasi-periodic orbits) can be seen scattered about the region. The extent of their sizes might well depend on the initial phase angle of Chariklo. It is noteworthy that these islands lie in the same rectangle which contains nearly all of the most long-lived orbits seen in the lifetime map. 

The orbits with relatively longer lifetimes and lower chaos are said to display stable chaos, and Figure~\ref{Chariklo_lifetime_MEGNO} shows that Chariklo has one of these orbits. 

\section{CONCLUSIONS}

The dynamical history of Chariklo and its rings was determined using the technique of numerical integration of massless clones backwards in time for 1 Gyr and by recording close encounters between test particles and giant planets.

We find that Chariklo most likely originated in an orbit beyond Neptune and was likely captured into the Centaur population via perturbations from Neptune sometime within the last 20 Myr. The backwards half-life with respect to removal of clones from the Centaur region is $\sim$ 3 Myr which is in good agreement with previous work on the backwards half-life of Chariklo with respect to removal from the Solar system.

Our results show that a small fraction of the clones of Chariklo spent some time significantly closer to the Sun than its current orbit. This suggests that it is possible but unlikely that Chariklo has undergone periods of cometary activity in its past - a result that mirrors the findings of \citet{HornerJ:2004a} that Centaurs can experience multiple periods of cometary behaviour throughout their lifetimes.


The critical distances of the Hill radius, tidal disruption distance, `ring limit` (defined as ten times the tidal disruption distance) and Roche limit were used to create a severity scale for close encounters based solely on the minimum distance obtained between the test particle and planet during the encounter.

More than 99\% of all close encounters over the course of our simulations were sufficiently distant that no impact on the structure of Chariklo's rings would be expected. Indeed, just 35\% of all clones experience an encounter within ring limit with one or other of the giant planets. In other words, 65\% of clones never experience a sufficiently close encounter to significantly disrupt the ring system in a single pass. We conclude that planetary encounters have likely not played a major role in influencing the structure of the rings.

Close encounters in which the test particle crossed the Roche limit were extremely rare - making up just 0.0034\% of the total sample of encounters observed. As result, we consider that it is highly unlikely that Chariklo's rings were created as a result of tidal disruption during such an encounter.

There is only a small chance that the rings have been replenished due to cometary activity in the inner Solar system. 


The lifetime of Chariklo-like orbits (orbits with different $a$ and $e$) are are found to be dependent on the eccentricity of the orbit with a general trend that orbits with higher eccentricities have shorter lifetimes. The crossing of Saturn's orbit plays a strong role in reducing the lifetime of an orbit.

Nearly all Chariklo-like orbits in the region bounded by $14\textnormal{ au}\le a \le 19\textnormal{ au}$ and $0\le e \le 0.8$ are strongly chaotic with only relatively small islands in $a-e$ space which are less chaotic. Chariklo's orbit ($a=15.8$ au, $e=0.172$) displays stable chaos by having a very chaotic orbit in a region with a relatively longer lifetime compared to nearby Chariklo-like orbits. 

Chariklo needs to be studied further to determine definitively if it shows evidence of past cometary activity. If it is ever proven that Chariklo was once active, it would support the idea presented here that in the past Chariklo could have had an orbit closer to the Sun where its rings could have been replenished.

\section*{ACKNOWLEDGMENTS}
We wish to thank the referee for their feedback, which helped to improve the flow and clarity of our work. This research has made use of NASA's Astrophysics Data System, NASA's JPL Horizons' database and the Asteroids Dynamic Site. TCH acknowledges support from KASI grant \#2015-1-850-04 and 2016-1-832-01. Numerical computations were partly carried out using the KASI/POLARIS and Armagh/ICHEC computing clusters.

\end{document}